\documentclass[runningheads]{llncs}
\usepackage{hyperref}
\usepackage{float}
\usepackage[T1]{fontenc}
\usepackage{caption}
\usepackage{graphicx}
\usepackage{etoolbox} 
\AtBeginEnvironment{thebibliography}{\tiny}

\begin{document}
\title{The Configuration of Space: Probing the Way Social Interaction and Perception are Affected by Task-specific Spatial Representations in Online Video Communication}
\titlerunning{The Configuration of Space}
\author{Yihuan CHEN\inst{1}\orcidID{0009-0004-2723-7398}
\and
Kexue FU\inst{1}\orcidID{0000-0002-2929-2663}
\and
Qianyi CHEN\inst{2}\orcidID{0009-0004-3768-6269}
\and
Zhicong LU\inst{3}\orcidID{0000-0002-7761-6351}
\and
RAY LC\inst{1}\orcidID{0000-0001-7310-8790}\thanks{Correspondences should be addressed to LC@raylc.org.}
}
\authorrunning{Y. Chen et al.}
\institute{
    Studio for Narrative Spaces, City University of Hong Kong, Hong Kong, China \\
    \email{\{yihuachen6-c, kexuefu2-c\}@my.cityu.edu.hk, 
           LC@raylc.org} \and
    College of Environmental Design, University of California, Berkeley CA, USA \\
    \email{qianyi.chen@berkeley.edu} \and
    College of Engineering and Computing, George Mason University, Fairfax VA, USA\\
    \email{calebluzc08@gmail.com}
}

\maketitle  % typeset the header of the contribution

\begin{abstract}
Humans live and act in 3D space, but often work and communicate on 2D surfaces. The prevalence of online communication on 2D screens raises the issue of whether human spatial configuration affects our capabilities, social perception, and behaviors when interacting with others in 2D video chat. How do factors like location, setting, and context subtly shape our online communication, particularly in scenarios such as social support and topic-based discussions? Using Ohyay.co as a platform, we compared a normal gallery interface with a scene-based Room-type interface where participants are located in circular arrangement on screen in a social support task, and found that participants allocated attention to the group as a whole, and had pronounced self-awareness in the Room format. We then chose a two-sided topic for discussion in the Gallery interface and the Room interface where participants on each team face-off against each other, and found that they utilized spatial references to orient their allegiances, expressing greater engagement with those farther away in digital space and greater empathy with those closer, in the Room over the Gallery format. We found spatial effects in the way participants hide from the spotlight, in perspective-taking, and in their use of expressive gestures in time on the screen. This work highlights the need for considering spatial configuration in 2D in the design of collaborative communication systems to optimize for psychological needs for particular tasks.

\end{abstract}
\section{Introduction}
In face-to-face communication, space provides crucial environment cues like object layout and seating arrangements, influencing how people perceive and communicate with each other \cite{russell_seating_1980}, and proxemics, which determines interpersonal communication based on the distances between individuals \cite{hall1966hidden}. When transitioning from 3D physical space to 2D screens in computer-mediated communication (CMC), the influence of space and social dynamics changes. Video in CMC offers limited environment cues, and there are different views on how non-verbal cues are perceived in CMC compared to physical space, with some arguing they're filtered out and others stating they can convey social information. Since the 2010s, video chat platforms like Skype have expanded, and the COVID-19 pandemic made video communication via platforms like Zoom integral to daily life, especially for remote work, learning, and maintaining relationships \cite{heshmat_family_2021}. In the current home-based remote meeting context, issues such as lack of non-verbals, insufficient natural spatial cues\cite{underhill2003experimental}, and cognitive loads\cite{bailenson_nonverbal_2021} have emerged, driving advancements in video calling user experience.

Many collaborative and communicative environments use notions of “space” and spatial organization to facilitate and structure interaction. Spatial organization can aid in mirroring real-life interactions, organizing activities and groups efficiently with better performance and less effort~\cite{tang2023perspectives}, and enhancing user engagement through immersive spatial metaphors~\cite{dourish_re-space-ing_2006}. Previous research on hybrid video communication shows that spatial organization, such as seating arrangements and perspectives, significantly impacts the level of unity, presence, and overall effectiveness of communication. 
It's crucial to recognize the impact of spatial configuration in 2D video chats during long-distance, multi-person remote communications and discussions despite the prevalence of hybrid meetings in professional settings. Understanding how this virtual spatial organization influences our interaction abilities, social perceptions, and behaviors is key to optimizing these digital encounters.

\begin{figure}
\includegraphics[width=\textwidth]{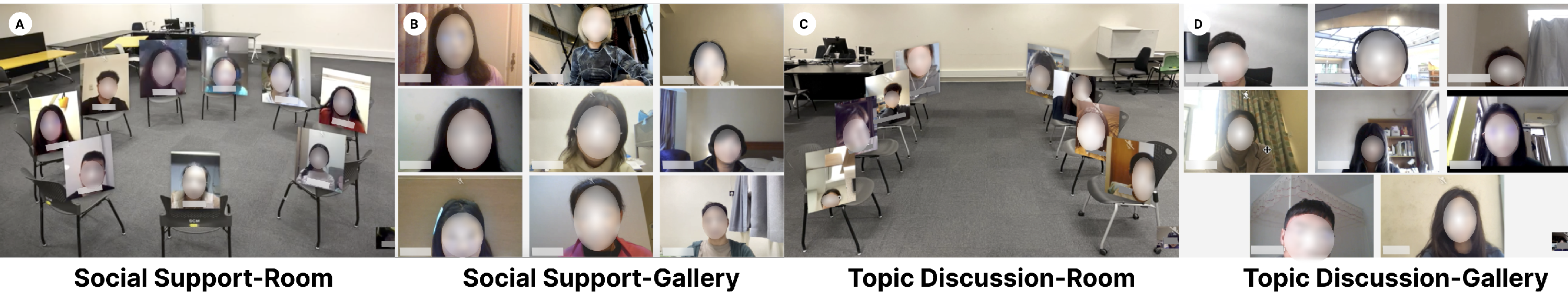}
\caption{Study setup for four conditions}
\label{fig:Overalltest}
\end{figure}

To understand how the spatial organization in online video communication influences social interaction and draw design implications for video calling UX, we conducted a qualitative study (Figure \ref{fig:Overalltest})  to investigate how the spatial configuration of a video chat space influences people’s perception and behaviors with others engaged in different tasks. Every group includes two conditions – a social support group and a topic discussion group. We set people in Room condition and normal Gallery condition with an equal number of people. The difference between Room and Gallery is the layout of the video frame and the chat room’s background: in Gallery condition, we use a grid layout displaying videos in equal size on a plain background, which simulates the "Gallery" view mode widely used in video conferencing platforms like Zoom; in Room condition, we use a classroom with chairs in it and the video frame “sit” in each chair. For the Social Support, the background of the Room condition is a classroom with a circle chair arrangement; for the Topic Discussion, the background of the Room condition is a classroom with an angular chair arrangement.

% Our data analysis from the video and interview revealed  satisfied engagement and empathy levels in the Room condition shown by participants. Our work also found that space can provide references for participants in visual search and identify their allegiances in the Room condition. In addition, this environment cue helped enhance team awareness for participants in the Room with a topic discussion task. Some other patterns and behaviors observed from the study are also discussed, including an intention to hide from the spotlight and hand gestures using spatial information. Besides, this work highlights the design implications of 2D video chat space and considerations for the spatial configuration in satisfying cognitive needs for tasks with different purposes. Overall, our work contributes to a better understanding of how space configuration in online video communication influences the interaction visually and conceptually under different contexts.
Our data analysis revealed satisfied engagement and empathy levels in the Room condition shown by participants. Our work also found that space can provide references for participants in visual search and identify their allegiances in the Room condition. In addition, this environment cue enhanced team awareness for participants in the Room with a topic discussion task. Some other patterns and behaviors observed from the study are also discussed, including an intention to hide from the spotlight and hand gestures using spatial information. Besides, this work highlights the design implications of 2D video chat space and the spatial configuration in satisfying cognitive needs for tasks with different purposes.

%Overall, our work contributes to how space configuration in online video communication influences the interaction visually and conceptually under different contexts.

\section{Related Work}

\subsection{Evolution of Spatial Cues in Video Communication Tools
}
% Early studies on the affordances of media space show that the "space" in video communication tools mediated by technology is significantly different from the physical space \cite{gaver_affordances_1992}. Harrison and Dourish's work visited the notion of "space" on a more conceptual level by proposing a distinction between "place" \cite{harrison_re-place-ing_1996}. In their work, “space” is described as geometrical arrangements that might structure, constrain, and enable certain forms of movement and interaction, while “place” denotes the ways in which settings acquire recognizable and persistent social meaning in the course of interaction~\cite{dourish_re-space-ing_2006}. 

Early studies on the affordance of media space show distinctions between "space" in video communication tools and physical space \cite{gaver_affordances_1992}. Harrison and Dourish's work conceptually distinguished "space" from "place", where “space” refers to geometrical arrangements that might structure, constrain, and enable certain forms of interaction, while “place” describes how settings acquire recognizable social meaning during interaction~\cite{harrison_re-place-ing_1996}. 
 
 % The benefits of physical space, which are crucial for interactions such as conversation initiation and object identification, are also useful in 2D communication for virtual collaboration. Spatial elements like identity, orientation, and activity focus improve design in these collaborative systems. They act as metaphors and replicate real-world behaviors, supporting natural human interactions in collaborative systems through spatial organization like fostering spatially guided behaviors~\cite{harrison_re-place-ing_1996}.  The 'spatial metaphor' is an important concept in HCI, which is defined as a conceptual tool that employs our understanding and perception of physical space to assist users in interacting with and navigating digital environments~\cite{wang2000study,alty1998metaphor}.

The benefits of physical space also enhance 2D virtual collaboration. Spatial elements such as identity, orientation, and focus on activity improve collaborative system design by replicating real-world behaviors and supporting natural human interactions. These elements act as spatial metaphors, defined as conceptual tools that employ our understanding and perception of physical space to help users interact and navigate in digital environments, affecting how people present themselves online~\cite{zhang_image_2025}.

% Previous research has investigated several systems that have consistent spatial arrangements. 
% Harrison and Dourish's work analyzes the MUDS system from the perspective of the spatial metaphor. MUDs~\cite{curtis1994muds} create virtual worlds divided into distinct locations or 'rooms,' enabling users to navigate and selectively engage in various events, activities, and conversations, fully integrating the spatial metaphor in their model of action and interaction~\cite{harrison_re-place-ing_1996}. 
% MAJIC~\cite{okada1994multiparty} represents a line of research prototypes that looked to create a virtual sense of shared space among remote collaborators. It displayed life-size participant images on a curved screen, enabling realistic eye contact and group interaction, but it was bulky and had visual flaws. 
% HERMES~\cite{inoue1997integration}, an innovative hybrid meeting prototype, blended remote and on-site participants by arranging video displays in a circular table layout. This design, however, shifted the focus of remote attendees' view to the active speaker. Despite its intent to improve interaction beyond standard flat-screen setups, remote participants faced a different UX and failed to benefit from the HERMES layout. This reveals the need for a more inclusive design catering to remote and in-person attendees equally.
Previous research has investigated several systems that have consistent spatial arrangements. 
Harrison and Dourish's work analyzes the MUDS system from the perspective of the spatial metaphor. MUDs~\cite{curtis1994muds} create virtual worlds divided into distinct locations or 'rooms,' enabling users to navigate and selectively engage in various events, activities, and conversations~\cite{harrison_re-place-ing_1996}. 
% MAJIC~\cite{okada1994multiparty} represents a line of research prototypes that display life-sized participant images on a curved screen, enabling realistic eye contact and group interaction, but was bulky and had visual flaws. 
HERMES~\cite{inoue1997integration}, an innovative hybrid meeting prototype, blended remote and on-site participants by arranging video displays in a circular table layout. However, this design shifted the focus of remote attendees' view to the active speaker. 
% Despite its intent to improve interaction beyond standard flat-screen setups, remote participants faced a different UX and failed to benefit from the HERMES layout. This reveals the need for a more inclusive design catering to remote and in-person attendees equally.

Spatial configuration in video chat has been developed in online platforms. For example, one group utilized 360 photos to enhance contextual awareness \cite{guo_as_2019}, and studied the role of video feedback in video-focused and content-focused layouts \cite{k_miller_meeting_2021}. Some features in recent video conferencing tools, such as virtual background, are also studied in terms of how they might influence person's perception \cite{Hwang_2021}. In addition, several new videoconferencing tools emerged during the pandemic. Ohyay.co \cite{ohyay_nodate} is an online platform that allows users to build different video chat rooms and Gather. town, another online platform with video chat \cite{gather_nodate}. On virtual platforms, mirrors can have a spatial effect \cite{fu_i_2023}, allowing people to be distracted by their avatars \cite{zhang_becoming_2025}.

% In summary, although many previous video chat systems have utilized spatial configuration, the visual design still has certain limitations, such as not being inclusive enough. Therefore, we aim to propose a more inclusive and equitable design for remote participants. We used ohyay.co as our platform to build two different layouts (Gallery-type and Room-type) and did testing.

\subsection{Spatial Configuration and User Interaction}

Later works, therefore not only view space on its physical attributes but also as a product of social practices \cite{dourish_re-space-ing_2006,brewer_storied_2008}. 
% Many previous studies have investigated how to improve the configuration of videoconferencing in order to make the social experience as similar to face-to-face meetings as possible. 
Previous studies have investigated how to improve videoconferencing configuration to gain a similar social experience as face-to-face meetings. Approaches taken included achieving mutual gaze \cite{okada1994multiparty}, capturing spatial cues \cite{buxton1992telepresence},and enhancing spatial information using multiple cameras with a spatially distorted screen \cite{nguyen_multiview_2007,ohara_blended_2011}.  

% The seating arrangement has been a focus in previous work  \cite{russell_seating_1980, mehrabian_seating_1971}. Studies show that different shapes of seating arrangement can result in different dynamics in the classroom. For example, changing from a row-based arrangement to a circle-shape arrangement brought an increase in classroom engagement \cite{falout_circular_2014-1}. People who sit in chairs arranged in a circular style reveal preferences on family-oriented content compared to angular shape \cite{zhu_exploring_2013}, while sitting in opposite rows is a common seating arrangement for debates \cite{minchen2007effects}. These studies suggest that in the physical space, the shape of the seating arrangement can be picked up by people and bring subtle influences on their perception and behaviors.  However, the impact and role of spatial configurations that either match or don't match the discussion topics are not thoroughly explored.
The seating arrangement has been a focus in previous work  \cite{russell_seating_1980}. Studies show that different shapes of seating arrangement can result in different social dynamics in the classroom. For example, changing from a row-based arrangement to a circle-shaped arrangement increased classroom engagement \cite{falout_circular_2014-1}. People who sit in chairs arranged in a circular style reveal preferences for family-oriented content compared to angular shapes \cite{zhu_exploring_2013}, while sitting in opposite rows is a common seating arrangement for debates \cite{minchen2007effects}. These studies suggest that in the physical space, people can pick up the shape of the seating arrangement, which can subtly influence their perceptions and behaviors. 
% However, the impact and role of spatial configurations that either match or don't match the discussion topics are not thoroughly explored.

In hybrid video-mediated communication, seating arrangements affect interaction frequency. Research~\cite{yamashita_impact_2008} found that distant participants seated side-by-side felt more united and reached better solutions than those seated across from each other. Another study on hybrid meetings showed that digitally composing everyone into a virtual room with unique, consistent viewpoints supports natural interactions, enhances co-presence, and reduces cognitive load~\cite{lc_sit_2024}. %Their discussions are mainly about hybrid meetings, focusing on how people pair up both offline and online. However, there's a gap in designing spatial configurations in 2D UX for more common multi-person online gatherings, like casual chats outside of work. Additionally, there's little emphasis on understanding and summarizing these settings from a behavioral pattern perspective.

%In sum, although previous work has explored different seating arrangements and how they influence interaction frequency in hybrid meetings, how spatial configurations impact human behaviors that match or do not match the discussion topics in 2D UX remain underexplored. Inspired by seating arrangements in real life, we designed two types of seating arrangements in the Room-type layout, one is circular style for the social support condition, and the other is two opposite rows for the topic discussion condition.

\subsection{Perception of Space in Computer-Mediated Communication}

% In a physical space, the perception of a social relationship is shaped by interpersonal distance. Hall's theory on proxemics suggests that the distance surrounding a person forms a space \cite{hall1966hidden}. Spaces with different distances are used for interactions with different levels of social closeness. For example, interaction at a close distance is for intimate friends, and a further distance is for socializing with acquaintances. This suggests that in a physical space, people may have different feelings when an individual interacts with them at a different distance, especially in real-world situations like exhibitions and performances\cite{lc_active_2023, lc_contradiction_2023, cao_dreamvr_2023}.

In a physical space, the perception of a social relationship is shaped by interpersonal distance. Hall's proxemics theory suggests that the distance surrounding a person forms a space \cite{hall1966hidden}. Spaces with different distances reflects different levels of social closeness. For example, closer interactive distance is for intimate friends, and a further distance is for socializing with acquaintances. 

% This suggests that in a physical space, varying interpersonal distances may evoke different feelings, especially in real-world situations like exhibitions and performances\cite{lc_active_2023,lc_contradiction_2023,cao_dreamvr_2023}.

Epstein's work on scene perception provides a possible explanation for understanding how human beings gather and process information from the environment and form their perception of space \cite{epstein_scene_2019}. According to this theory, when looking at an image depicting a scene, the observer extracts information from the scene properties and categorizes it on different levels. This includes low-level concrete properties (e.g., edges), mid-level elements (e.g., layout), and high-level abstract properties (e.g., category). This information can be used in many cognitive functions, including searching for an object in the scene and recognizing the context of the scene. 

% The perception of the virtual space of video chat can be different from perceiving the scene of an image. Firstly, in a video chat, each participant has a different background, which means a person in a video chat processes information from not only one scene but also multiple scenes simultaneously. This can bring more cognitive load for people in the video chat \cite{Hwang_2021}. Different spatial configurations of the video chat space also bring in different types of information for people to process. For example, in a Gallery-type layout where the video feeds of each participant are displayed together on a plain background, the environment cues mainly come from the background of participants. However, in a Room-style configuration with a background image presenting a scene in the shared space, the scene perception for people in the video chat becomes more complicated as they will process two different types of scenes. Besides, in the Room configuration, the participants are located in the scene in a way different from several people standing in a physical space.

The perception of the virtual space of the video chat can be different from perceiving the scene of an image. Firstly, in a video chat, each participant has a different background, which means a person in a video chat processes information from one scene and multiple scenes simultaneously. This can increase the cognitive load for people in the video chat \cite{Hwang_2021}. Different spatial configurations of the video chat space also affect information processing. For example, in a Gallery-type layout where the video feeds are displayed together on a plain background, the environment cues mainly come from the background of participants. However, in a Room-style configuration featuring a shared background space complicates perception as participants will process two different types of scenes. This has implication for remote engagement strategies that uses the background for storytelling\cite{lc_contradiction_2023,lc_active_2023,lc_presentation_2022}. 
% Besides, in the Room configuration, the participants are located in the scene differently from people standing in a physical space.

Thus we set out to discover how different spatial designs influence the social relationships between remote participants and find spatial configurations that allow a sense of co-presence while reducing cognitive load.
 
\section{Method}
% % Drawing from insights gained in prior research, we embarked on a prototyping process to delve into how video conference interfaces with diverse spatial configurations shape the user experience, particularly in multi-person online meetings. This endeavor includes examining the influence of various spatial layouts on discussions across different topics.
% Drawing insights from prior research, we embarked on a prototyping process on how video conference interfaces with diverse spatial configurations shape the user experience, particularly in multi-person online meetings. Our study focuses on two specific spatial configurations: Gallery-type and Room-type. The aim is to not only scrutinize the nuances in user behavior and interaction within these setups but also to derive design considerations for enhancing the user experience (UX) of video-calling prototypes. Key areas of interest are: (1) crafting an interaction space that is both inclusive and equitable for remote attendees, (2) leveraging spatial cues to integrate gaze awareness and gesture use seamlessly, (3) cultivating a sense of co-presence among participants, (4) aiming to reduce cognitive load during meetings, and (5) working towards alleviating anxiety in virtual communication contexts.

\subsection{Study Set-up}
\subsubsection{Participants} 
Participants (N=34) (Appendix, Table \ref{tab1}) were college students recruited through social media and posts on campus. Participants were divided into Social Support (N=18, 13 females, 5 males) and Topic Discussion (N=16, 12 females, 4 males). All participants signed an online consent form and were given \$15. They were also provided with an honorarium for completing the study. 
    
% The user study was carried out on ohyay.co,
\subsubsection{Study Design} The user study was conducted on ohyay.co, a versatile video platform that enables users to create custom video chat room layouts tailored to specific research requirements. Ohyay's UX design allows for the placement of video streams against a communal digital backdrop, such as a meeting room or lecture hall, with the layout varying based on the chosen scene. This platform empowers participants to personalize their experience by dragging and dropping video streams to preferred locations. Moreover, the uniformity of the scene viewed by all participants ensures a consistent visual experience, fostering a sense of cohesion in the meeting.

% In contrast, the Gallery-type, commonly found in numerous commercial video conferencing systems such as Microsoft Teams, Zoom, Google Meet, and Cisco Webex, presents a traditional grid view of video streams. This layout typically occupies the entire display without screen sharing, with each user having their own distinct space within the grid.

In our study, two different layouts were configured on Ohyay: Room-type and Gallery-type. The Room-type features spatial seating arrangements and a consistent background, replicating a classroom setting with chairs arranged either in a circle or on two sides according to tasks. In contrast, the Gallery-type, commonly found in numerous commercial video conferencing systems such as Microsoft Teams, Zoom, and Google Meet, presents a traditional grid view of video streams. This layout occupies the entire display without screen sharing, with each user in their own distinct space within the grid. To generalize our study of spatial configuration’s influence on the video chat, we designed two different tasks (Figure \ref{fig:Overalltest}) –Social Support, in which people sit together, share personal experiences and support each other; and Topic Discussion, in which people hold opposite opinions about one topic sit and discuss together. 

In the Social Support task, participants (N=18) discussed challenges and experiences during COVID-19 to support other participants. The prompt given by the researcher is: “The uncertainty of the pandemic has brought stressful and anxious moments to people all over the world. Today, we’re here to support everyone in combating the upset feelings toward this issue. Everything in this chat will be kept confidential.” All the participants were equally divided into two 9-person groups with two different spatial configurations. One group called Gallery is a grid layout for simulating the Gallery interface in existing online video platforms, which is typically used for multi-person remote meetings and displays the video of each participant in equal size. Another spatial configuration is a Room using a background image of a classroom, a photo taken by the researchers. In the photo, 9 chairs were placed in a circle, mimicking the circular seating style usually for family-oriented content discussion \cite{zhu_exploring_2013}. Each participant's video frame was positioned based on the location of each chair in the background. Based on the position, the perspective of the video frame is adjusted using the rotation and skew parameters on each axis. Except for the layout, all settings are the same in two conditions. This conversation for each group lasted around 30 minutes.

\begin{table}[H]
    \centering
    \caption{Summary demographics of the participants.}
    \label{tab1}
    \begin{tabular}{|l|l|l|l|l|l|l|l|}
    \hline
    \multicolumn{4}{|c|}{Social Support} & \multicolumn{4}{c|}{Topic Discussion} \\ \hline
    ID & Condition & Age group & Gender & ID & Condition & Age group & Gender \\ \hline
    P1  & Room      & 18$\sim$24 & F & P19 & Room      & 18$\sim$24 & F \\ \hline
    P2  & Room      & 18$\sim$24 & F & P20 & Room      & 18$\sim$24 & M \\ \hline
    P3  & Room      & 25$\sim$31 & M & P21 & Room      & 18$\sim$24 & M \\ \hline
    P4  & Room      & 18$\sim$24 & F & P22 & Room      & 18$\sim$24 & F \\ \hline
    P5  & Room      & 18$\sim$24 & F & P23 & Room      & 25$\sim$31 & F \\ \hline
    P6  & Room      & 18$\sim$24 & F & P24 & Room      & 18$\sim$24 & F \\ \hline
    P7  & Room      & 25$\sim$31 & M & P25 & Room      & 18$\sim$24 & F \\ \hline
    P8  & Room      & 18$\sim$24 & F & P26 & Room      & 18$\sim$24 & F \\ \hline
    P9  & Room      & 25$\sim$31 & M & P27 & Gallery   & 18$\sim$24 & M \\ \hline
    P10 & Gallery   & 25$\sim$31 & F & P28 & Gallery   & 18$\sim$24 & F \\ \hline
    P11 & Gallery   & 18$\sim$24 & F & P29 & Gallery   & 18$\sim$24 & F \\ \hline
    P12 & Gallery   & 25$\sim$31 & F & P30 & Gallery   & 18$\sim$24 & F \\ \hline
    P13 & Gallery   & 18$\sim$24 & F & P31 & Gallery   & 18$\sim$24 & F \\ \hline
    P14 & Gallery   & 18$\sim$24 & F & P32 & Gallery   & 18$\sim$24 & F \\ \hline
    P15 & Gallery   & 18$\sim$24 & M & P33 & Gallery   & 18$\sim$24 & M \\ \hline
    P16 & Gallery   & 18$\sim$24 & F & P34 & Gallery   & 18$\sim$24 & F \\ \hline
    P17 & Gallery   & 32$\sim$38 & F &  &  &  &  \\ \hline
    P18 & Gallery   & 18$\sim$24 & M &  &  &  &  \\ \hline
    \end{tabular}
    \end{table}

% The video frame of each participant was positioned based on the location of each chair in the background. Based on the position, the perspective of the video frame is adjusted using the rotation and skew parameters on each axis. Except for the layout of each video chat room, we kept the rest of the settings the same in two conditions. This conversation for each group lasts around 30 minutes.

% 16 people (N=16) participated in the Topic Discussion task. Based on their background, the task topic is about whether the university should launch a compulsive policy that undergraduate students need to finish a full-time internship in one term. All the participants were equally divided into two 8-person groups. One condition is also the normal Gallery condition. Another spatial configuration is Room, using a background image of a classroom. 
16 people (N=16) participated in the Topic Discussion task. Based on their background, the topic is whether the university should mandate a full-time internship in one term for undergraduate students. All the participants were equally divided into two 8-person groups. With one in normal Gallery condition, another group was in Room configuration with a background image of a classroom. In the photo, 8 chairs were placed in two opposite lines, applying the debate-style seating arrangement. Each participant’s video frame was positioned based on the chair locations in the background. Except for the layout of each video chat room, All other settings were consistent between the two conditions. This conversation for each group lasts around 30 minutes. For both Room and Gallery conditions, participants were randomly and equally assigned to opposing viewpoints and discussed their standpoints rationally.
% In the photo, 8 chairs were placed in two lines, applying the opposite-line seating style often used for debates. The video frame of each participant was positioned based on the location of each chair in the background.  Except for the layout of each video chat room, we kept the rest of the settings the same in two conditions. This conversation for each group lasts around 30 minutes. 
% participants were randomly and equally divided into two groups holding opposite views. They discussed their standpoint rationally. 
\subsubsection{Procedure}
Recruited participants were randomly assigned to different conditions, including 9 people in the Room with a social support task, 9 people in the Gallery with a social support task, 8 people in the Room with a topic discussion task, and 8 people in the Gallery with a topic discussion task. Two researchers monitored the study, one as the note-taker and one as the facilitator. In the beginning, one researcher facilitated a short self-introduction session for participants to get familiar with each other. Then, the researcher gave a topic for them to discuss. Participants in Room and Gallery conditions with the same task followed the same procedure. The researchers did not make interventions during the 30-minute discussion. The process of the video chat was recorded in video form for data analysis. At the end of the study, we also conducted post-study interviews. Details of the interview are elaborated on in a later session.

\subsection{Data Collection: Interview}

To garner a deeper understanding of the survey responses and trends identified in the data, follow-up interviews were conducted (N=34).  

The interviews were conducted in the participants’ native language less than 12 hours after the group chats to make sure they could remember and elaborate on the details of the video chat vividly. The interviews were conducted using video or voice calls. Each interview lasted about 30 minutes. The interviews were semi-structured, with questions probing participants’ experiences during the group chat. After recalling people’s feelings in that process, we specifically asked them questions related to “spatial perception”-- how they felt about the “space” in the video chat room and whether they were influenced by the spatial perception. Moreover, we asked what they thought about the gap between group chat online and in the physical space. With approval from interviewees, all interviews were audio-recorded and transcribed.

\subsection{Data Analysis: Video and Interview}

The video recordings and interview transcriptions were analyzed using an open coding method~\cite{khandkar2009open}. For interviews, researchers coded segments of the interview scripts by themselves, then came together to discuss their codes and reached an agreement for a codebook. Each author then categorized the rest of the transcriptions using the codebook. All the codes were translated into English.

% For video recordings, the code theme is based on the summary of interaction features that emerged in initial video analysis and also refers to methodologies for analyzing the group interactions in communication, such as the Discussion Coding System (DCS) developed by Schermuly \cite{schermuly_discussion_2012}.
For video recordings, the code themes were developed based on initial interaction features and communication methodologies like the Discussion Coding System (DCS) \cite{schermuly_discussion_2012}. Among the code themes, facial expressions -- smiles, measured people’s attitudes to the chat content and how the social dynamic changed in different conditions. The body language -- "handclap," “nod head,” and "look other places outside the screen", can evaluate the participants’ engagement in different conditions, which is related to the level of connection with others. Other body language -- "turn heads" and "point to someone" are usually used to locate others in the physical space, and we are supposed to measure whether people take actions related to their spatial perception. 

% This time length is neither too short to generate many empathy codes nor too long to miss many interaction details.
% The videos were analyzed in a time series. Researchers coded the video every 20 seconds. This interval was chosen to balance detailed interaction capture without overwhelming data. Participants’ performances were coded individually due to varying experiences and responses in the chat room. At last, we plotted all the events of everyone and summed events across conditions. 

% \section{Video Data Results}
% % The video data allowed us to dissect how each session occurred in terms of the way people talked to each other, and in terms of the nonverbal gestures that people made with each other.

% Our analysis of participant speech patterns over time in four meetings revealed no distinct behavior patterns across conditions. However, a noticeable trend emerged in the Room-Social Support setting at the beginning of the meetings, where participants tended to speak in a clockwise sequence around the round table. This pattern was not as evident in the Gallery condition or other meeting scenarios (refer to Appendix, Figure \ref{fig:Videospeaking}).

\section{Qualitative Results}
%  % we aim to investigate how the
% As discussed in the method session, we investigate how the spatial configuration influences the social dynamics of a group chat under different tasks. Our analysis reveals patterns around how the participants gathered and processed the spatial information using visual cues, and how they formed the perception of self and others in group video chats with different spatial configurations.

\subsection{Gathering and processing spatial information using visual cues}
% we concluded emerging themes about how the participants picked up these visual cues from both the conversation and the video chat environment, and how it related to their behaviors.

Our study investigated information collecting and processing in the video chat. From the interview analysis, we concluded emerging themes about how the participants picked up these visual cues from conversation and environment, and how the cues related to their behaviors. There were two main aspects when a participant gathered information through observations: tagging the participants with a spatial location and glancing through the group in a certain order.

\subsubsection{Tagging the participants with a spatial location.}

One theme was searching and locating other individuals in the group video chat. In the real world, a scene can provide useful guidance for visual search with its properties \cite{epstein_scene_2019}. Our analysis found similar behaviors happening in the virtual space of the group video chat under the Room condition - participants could locate other participants with a virtual “seat” and glance in a certain order using the chairs in the background image. 

Instead of finding different objects in a scene, the “objects” of the video chat were the videos of participants. The need to locate a specific participant during the group video chat occurred more often in the topic discussion task where participants needed to identify who was on their or the other side. In the Room layout, by putting the pro and con team separately on the left and right side of the room, participants were able to “label” other participants based on their standpoint, thus saving the time of repeating this process, e.g. \emph{“It’s easier to locate other participants using this position in the space so that I don’t need to find them every time” (P26, Room - Topic Discussion)}. 
% This was considered an advantage of Room compared with Gallery, where every participant was given a random position. Participants in the Gallery setting, who were engaged in the same topic discussion task, reported exerting mental effort in identifying their team members. \emph{"One boy seems confused about which side he was on, so I have no impression of him at the beginning. Yes, I was also confused. I thought he was on the opposite side, so I didn't pay attention to his talking" (P27, Gallery - Topic Discussion)}. 

% Therefore, the use of chairs in the scene attaches a spatial attribute for each participant, which can be used by other participants in memorizing and referring to an individual,  such as \emph{“the girl opposite to me” (P23, Room - Topic Discussion).} The way of referencing an individual in Room is similar to the way we use in a physical space. In the Gallery setting, it was observed that individuals were seldom referred to by their "position" (e.g. the person sitting opposite me). It may be because the display of one’s video was randomized, and on each participant’s screen, the order of the video display may not be the same.
This was an advantage of Room over Gallery, where participants were randomly positioned. In the Gallery setting, participants were engaged in the same topic discussion task, reported exerting mental effort in identifying their team members. \emph{"One boy seems confused about which side he was on, so I have no impression of him at the beginning. Yes, I was also confused. I thought he was on the opposite side, so I didn't pay attention to his talking" (P27, Gallery - Topic Discussion)}. Therefore, the use of chairs attaches a spatial attribute for each participant, which can be used by other participants in memorizing and referring to an individual, such as \emph{“the girl opposite to me” (P23, Room - Topic Discussion).} In the Gallery setting, it was observed that individuals were seldom referred to by their "position" (e.g. the person sitting opposite me). It may be because the video display was randomized, and the order of the display varied on each participant's screen.

% However, this visual search guidance also has its limitations. Although the chairs in the scene helped participants establish a connection between an individual and their location on the scene, this information did not help them process time-sensitive information in the group chat, such as which individual is currently talking in the group chat, e.g. \emph{“I think in Room, it is chaotic because I keep finding who is talking” (P21, Room - Topic Discussion).} 

However, this visual search guidance has its limitations. Although the chairs helped participants link individuals to their locations, this spatial information did not help them process time-sensitive information in the group chat, such as which individual is currently talking, e.g. \emph{“I think in Room, it is chaotic because I keep finding who is talking” (P21, Room - Topic Discussion).} Therefore, it seems that whether this visual guidance is considered useful depends on the extent to which an individual relies on visual information and whether they care more about short-term information, such as who is talking, or long-term information, such as the location. Some visual-focused participants may find the visual guidance useful, while some content-focused participants may choose to filter out the visual information, e.g. \emph{“I’ll choose not to look at the screen, and take notes which helped me be more focused [...] those visual things may be a distraction for me” (P34, Gallery - Topic Discussion)}, or memorize another individual based on their opinions rather than using the spatial information, e.g. \emph{“I mainly focus on their content, and then realize whether they have the same opinion as me or not [...] did not remember who is in the same group with me” (P32, Gallery - Topic Discussion).}  

\subsubsection{The order of a glance.} 

% In addition to using a chair as the spatial reference to locate one participant, the chairs as a whole also allowed participants to scan through the group members in sequential order. According to the scanning behavior reported by the participants during interviews, the scan order varied based on the specific visual cues in the virtual space. In the context of this study, for some participants, it was the arrangement of chairs in the background image (in a circle vs on opposite sides of a table) that influenced the way participants scanned through video frames in the video chat room. For instance, in the topic discussion Room where the video frames were positioned in two opposite rows, participants said they \emph{"scanned from closer to the screen to further from the screen (P26, Room - Topic Discussion)"}. 

In addition to locating one participant, the chairs as a whole also allowed participants to scan through the group members in sequential order. According to the scanning behavior reported by the participants, the scan order varied based on the specific visual cues. For some participants, it was the arrangement of chairs in the background image (in a circle vs on opposite sides of a table) that influenced the way participants scanned through video frames in the video chat room. For instance, in the topic discussion Room where the video frames were positioned in two opposite rows, participants said they \emph{"scanned from closer to the screen to further from the screen (P26, Room - Topic Discussion)"}. 

\begin{figure}
\includegraphics[width=\textwidth]{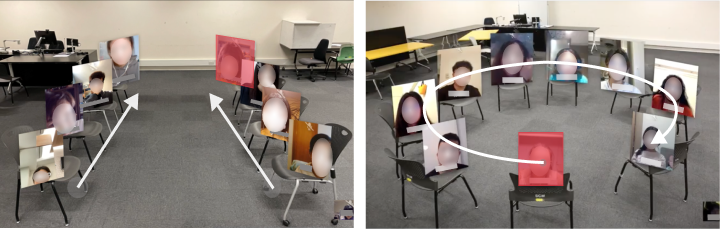}
\caption{Glance behaviors reported by Room participants. The participant who reported this glance behavior is marked with a red rectangle. The arrow illustrates the direction of the glance.}
\label{fig:glanceR}
\end{figure}

% While in the social support Room where chairs were arranged in a circle,
While in the social support Room with chairs in a circle, a participant “glanced others in the room in the clockwise order,” and another participant also “looked around” (P1, P9, Room - Social Support). This behavior in both conditions (circle vs line) is a one-way scan that had a start point, which was either the front and center seat of the circle or the front seat of a line that was closest to the screen (Figure \ref{fig:glanceR}). In both conditions, the computer screen was used as the frame of reference instead of the video of the participants. 

\begin{figure}
\includegraphics[width=\textwidth]{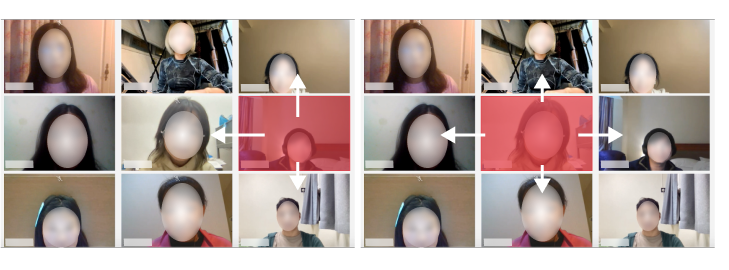}
\caption{Glance behaviors reported by Gallery participants. The participant who reported this glance behavior is marked with a red rectangle. The arrow illustrates the direction of the glance.}
\label{fig:glanceZ}
\end{figure}

However, a few Gallery participants (P14, P15, Gallery - Social Support; P31, Gallery - Topic Discussion) noted that they glanced through the video around their own video, such as allocating the attention “mainly based on a circle using my video as the center” (P15, Gallery - Social Support) (Figure \ref{fig:glanceZ}). This was reported by people whose videos were on different locations of the grid layout in Gallery condition, including the center and the right side on the second row. This may indicate two different types of viewpoints adopted by participants during the video chat.
% , which we’ll elaborate on further in the next section. 

\begin{figure}
    \centering
    \includegraphics[width=0.7\textwidth]{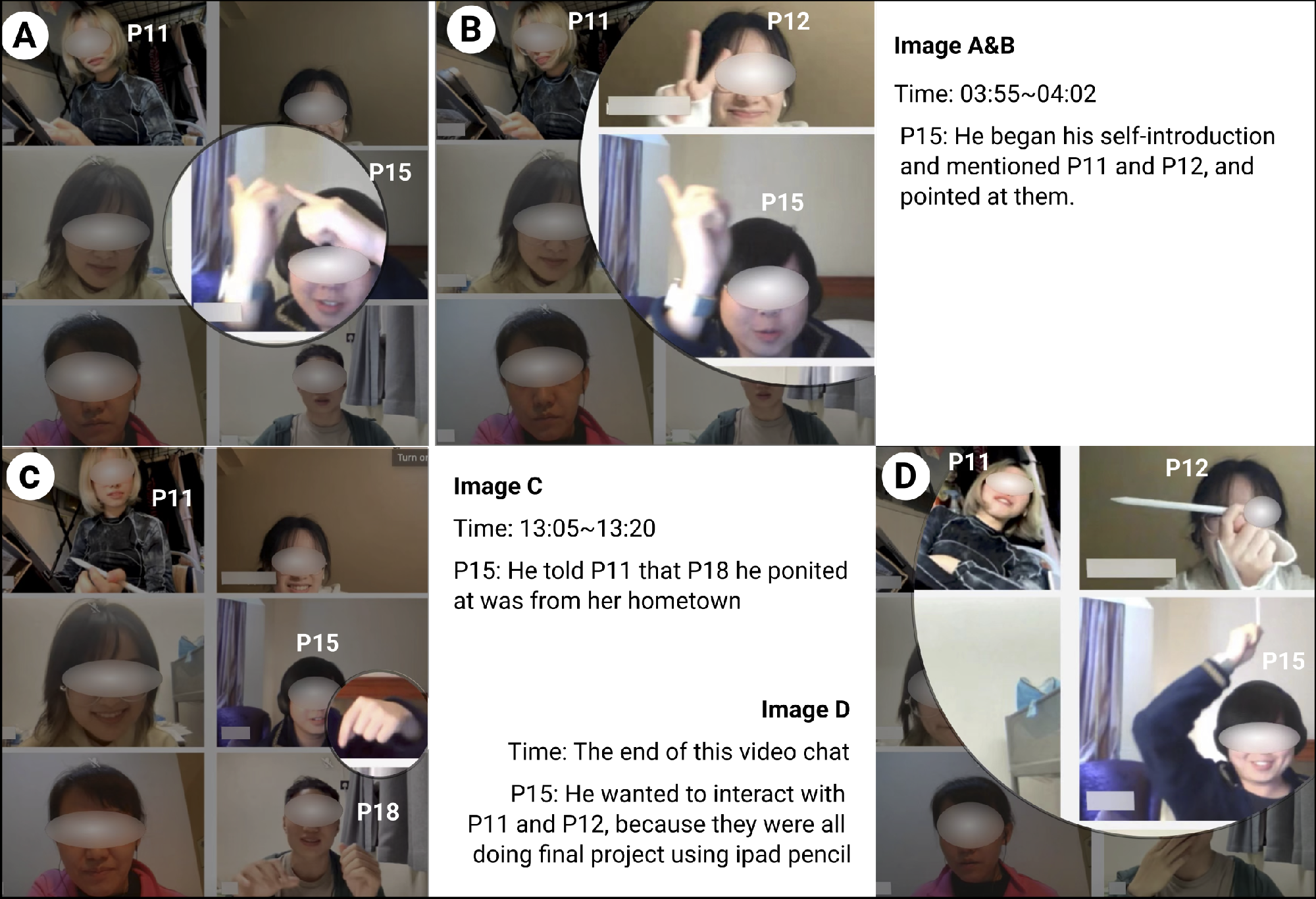}\caption{Social Support - gallery: Participant utilized spatial information to interact with others.}
    \label{fig:findingpoint}
    \end{figure}

\subsubsection{Spatial-based non-verbal communication}
\begin{figure}
\centering
\includegraphics[width=0.8\textwidth]{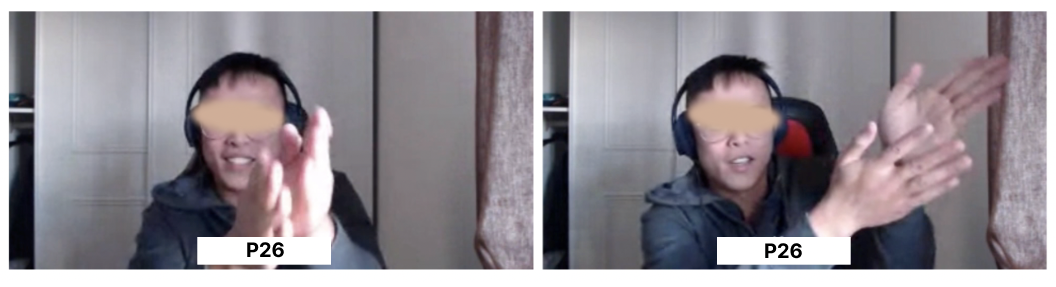}  
\caption{In the follow-up interview, a participant showed how he used body language "turn to the next " or "point a specific direction" to let people continue the topic.}
\label{fig:bodylanguage}
\end{figure}

Spatial cues enhanced nonverbal communication in both Room and Gallery conditions. In Room-Topic discussion, participants used directional gestures like turning or pointing to manage turn-taking (Figure \ref{fig:bodylanguage}). Similarly, in Gallery-Social Support, participants utilized gestures toward nearby videos, such as pointing to someone positioned above them, which effectively warmed up the atmosphere and elicited positive responses (Figure \ref{fig:findingpoint}).

Spatial proximity influenced behavioral synchronization. Participants sitting nearby tend to exhibite similar patterns - laughing together, mirroring body language, and coordinating microphone usage. As P19 noted, \emph{"When the person next to me was sharing a distressing experience in COVID-19 because they were 'sitting' beside me, I felt the urge to tap my partner on the shoulder, but then I realized that it might not be perceptible."} This suggests participants leveraged spatial cues in the 2D environment to strengthen interpersonal connections.

\subsubsection{Perception of others: Sense of belonging and connection shaped by spatial arrangement}
Room settings distinctly affected participants' sense of belonging across different tasks. In Topic Discussion, spatial arrangement tends to reinforce team identity, as P24 described: \emph{"It gave me a strong sense of debating and confrontation since the layout puts two parts separately and makes the boundary obvious. For example, I feel the person sitting opposite me and I are on different teams, and people sitting on the same side with me are on my team"}. This spatial division enhanced team awareness but could sometimes feel overly confrontational.

However, when participants' opinions diverged from their spatial positioning, the effect diminished. This mismatch between spatial cues and actual behavior highlighted the limitations of static spatial arrangements. As P24 noted, \emph{"Since the goal is not to rather than fighting against the people on the other side, I'm not sure whether this layout is too confrontational for a group discussion"}.

In Social Support sessions, circular arrangement fostered inclusivity. P4 described feeling "in a real environment with classmates," while P7 observed that \emph{"Under the Room-type setting, the feeling that everyone was in the same environment reduced distractions, positioning the Room-type as an intermediary state between traditional video conference interfaces and real-life interactions."} The shared classroom background and circular setup enhanced group cohesion and mutual support.

\begin{figure}
\centering
    \includegraphics[width=0.8\textwidth]{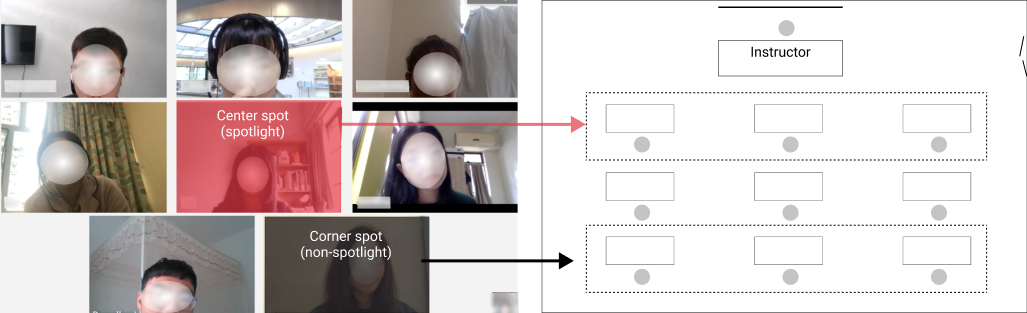}
    \caption{A spatial metaphor was made by the participants: "Sitting at the corner of the Gallery is like sitting at the back rows of the classroom."}
    \label{fig:classroom}
    \end{figure}

\subsection{Forming the Perception of Self, Others, and the Space}

\subsubsection{Perception of self: “Hiding from the spotlight” in the video chat as sitting in the back rows of the classroom}

A theme that emerged from the analysis was an intention to “hide from the spotlight” or avoid receiving too much attention from others during the video chat. This point was raised by participants from both Gallery and Room conditions. A participant in the Room condition highlighted feeling less anxious than Zoom's speaker view, where the person speaking would be prominently magnified, expressing that
\emph{“In my previous experience of using Zoom, when someone is talking, their video will be put in the middle with a larger size, and everyone knows who you are and looks at you. From that point, I feel more comfortable on this platform” (P23, Room - Topic Discussion)}. 

Another Room participant noted a different feeling compared with previous experience in Zoom,
\emph{“I feel more awkward in Room because everyone's video is displayed in a rectangle shape, and sometimes I will look at every one of the videos, which makes me feel stressed. [...] It's like when having face-to-face communication, I feel a lot of people are watching me, which made me feel uncomfortable” (P4, Room - Social Support)}

In addition to observations that the Room setting alleviated feelings of nervousness, feedback from a Gallery participant indicated that the sensation of being "in the spotlight" contributed to video chat anxiety. The desire to "stay out of the spotlight" could help lessen this anxiety. As noted by some participants, this tendency to avoid attention in a video chat was akin to choosing a seat in the back rows of a physical classroom (refer to Appendix, Figure \ref{fig:classroom}). It's the perceived focus from other participants that heightened this anxiety. For instance, a participant positioned in the bottom right corner of the Gallery grid layout expressed that,\emph{“This corner spot brought me a safe and comfortable feeling just like sitting at the back row of a classroom [...] probably because it does not receive much attention from other people [...] As long as I’m not in the center spot, which is like the first row of a classroom” (P34, Gallery - Topic Discussion).} 
% \end{quote}% \begin{quote}

% \setlength{\partopsep}{0pt}
% This observation highlighted a correlation between anxiety levels and the amount of attention a participant perceived they were receiving in a video chat
This observation highlighted a correlation between anxiety levels and perceived attention in video chats. Seats that draw less attention tended to be associated with lower anxiety, whereas those under greater scrutiny from others could lead to increased nervousness. Consequently, positions within the video chat environment could be categorized as "spotlight" or "non-spotlight" based on the perceived level of attention. The definition of a "spotlight" seat varies; it could be visually prominent (like the central position in a Gallery grid, often perceived as the most attention-receiving spot) or adjacent to an individual playing a special role (such as the host of an event or the lecturer in a class). Each participant's perception of attention could differ, influencing their sense of being "in the spotlight."

Furthermore, participants felt less anxious when observing others' facial expressions in video chats compared to face-to-face interactions, attributing this to a sense of "hiding." This feeling is linked to the attention perceived by others. In contrast to direct face-to-face communication, where observing someone's facial expression is a conspicuous action, online video chats made such observations less visible, reducing the awkwardness associated with being watched. For instance, one participant noted, \emph{"In a physical space, direct eye contact often feels awkward. But in the Room setting of a group video chat, I can observe everyone without them knowing, which alleviates the pressure of direct interaction" (P1, Room - Social Support).} Another participant from a Gallery discussion observed, \emph{"In face-to-face communication, scrutinizing everyone's facial expressions can seem impolite. In online discussions, however, you can observe not just the speaker but also the reactions of others without the discomfort of prolonged direct eye contact" (P32, Gallery - Topic Discussion).} This preference for less direct eye contact in video chats, as another participant explained, stemmed from the potential awkwardness of being observed too intently in a face-to-face setting (P16, Gallery - Social Support). In summary, the reduced visibility of observation behaviors and the avoidance of direct eye contact in video chats contributed to a sense of "hiding," which could alleviate social anxiety.

\section{Discussion}
% In our study, we explored two prototype interfaces, the Gallery-type and the Room-type in two discussion scenarios: Topic Discussion and Social Support. Our focus was to understand how these interfaces and contexts influence user experience in virtual settings.

% % In our study, we explored two prototype interfaces, the Gallery-type and the Room-type, in the context of two discussion scenarios: Topic Discussion and Social Support. Our focus was to understand how these interfaces and contexts influence user experience in virtual settings.

% % The Room-type interface is particularly noteworthy for its unique features. Firstly,
% Firstly, the Room-type interface provides enhanced spatial cues through its chair arrangement, creating a two-dimensional level of "space". This spatial arrangement allows for a more engaging and realistic remote communication experience. Secondly, the background image of Room-type offers a consistent spatial context for participants, essential for forming a "place" in a 2D video conferencing interface\cite{harrison_re-place-ing_1996}.

\subsection{Adaptation from Gallery-type to Room-type: Forming “Space” and Enhancing Visual Cues}

The Room-type interface embodies three key spatial features from a spatial model for collaboration\cite{harrison_re-place-ing_1996}: \textit{Relational Orientation and Reciprocity}, \textit{Partitioning}, and \textit{Proximity and Action}. These features represent aspects of "space" that significantly impact the user experience in virtual settings.

\textit{Relational Orientation and Reciprocity} emphasizes spatial references and orientations to establish context. In Room-type settings, the circular chair arrangement aids in creating a sense of order, particularly beneficial for tasks requiring sequential interaction like self-introductions or games like Mafia (players need to debate one by one and vote). This visual cue of a circle, either clockwise or anti-clockwise, is in contrast to the Gallery layout's grid structure. The latter lacks intuitive order, leading to random participation sequences in tasks. Furthermore, Room-type's spatially informed seating enhances nonverbal communication. Participants use gestures directed by spatial layout for interaction, making it closer to real-world scenarios.

% Secondly, \textit{Partitioning} plays a crucial role in spatial dynamics, using distance to define activity and interaction levels, akin to the 'boundary' concept highlighted by participants. This demarcation fosters team identity, as noted in Epstein's scene perception framework \cite{epstein_scene_2019}, where background elements like chairs contribute to cognitive functions like team recognition. Contrasting with Gallery-type interfaces lacking such spatial clarity, this approach reduces confusion in team identification.

Secondly, \textit{Partitioning} is crucial in spatial dynamics, using distance to define activity and interaction levels, akin to the 'boundary' concept highlighted by participants. This demarcation fosters team identity, as noted in Epstein's scene perception framework \cite{epstein_scene_2019}, where background elements like chairs contribute to cognitive function. Contrasting with Gallery-type interfaces lacking such spatial clarity, this approach reduces confusion in team identification. Room-type video chats, through specific seating arrangements, establish zones that enhance immersion. Circular seating in Social Support scenarios, for example, promotes engagement, whereas a dual-column layout in Topic discussions strengthens team cohesion. This suggests that 2D spatial layouts can influence user perception and behavior. But the discrepancies between anticipated and actual layouts in video chats may counteract these advantages by increasing cognitive load\cite{sweller1998cognitive}.

 % Room-type video chats, through specific seating arrangements, establish zones that enhance immersion in tasks such as Topic discussions and Social support. Circular seating in Social support scenarios, for example, promotes engagement, whereas a dual-column layout in Topic discussions strengthens team cohesion. This suggests that 2D spatial layouts can reflect real-life spatial interactions, influencing user perceptions and behaviors \cite{sailer2021differential, mehrabian_seating_1971}. Nonetheless, discrepancies between anticipated and actual layouts in video chats may counteract these advantages by increasing cognitive load, as indicated in prior research \cite{sweller1998cognitive}.

\textit{Proximity and Action} highlight the impact of spatial proximity on interactions. Participants in Room-type reported feeling more connected to those in close proximity, mirroring real-life interactions. This spatial property enhances the engagement in virtual meetings.

In summary, the spatial layout of the Room-type video interface, with its scene-specific arrangement, plays a crucial role in enhancing user experience in online meetings. Future research should explore the impact of different object layouts as environmental cues. For instance, investigating how various seating arrangements like squares, rows, circles, or columns influence user interaction would be insightful. Techniques like eye-tracking could provide quantitative data on these interactions \cite{jacob2003eye}. Additionally, comparing the spatial effects of different seating formations within the same task, such as a topic discussion in a Room setting with varying chair arrangements, would further our understanding of spatial dynamics in virtual meetings.

% Harrison and Steve (1996) highlight the shift from space to place, emphasizing that our behaviors are influenced more by the sense of place than space, as seen in different activities in conference halls versus theaters \cite{harrison_re-place-ing_1996}. 
\subsection{Adaptation from Gallery-type to Room-type: Forming “Place” and Enhancing Ambiance}
Harrison and Steve (1996) highlight the shift from space to place, emphasizing that our behaviors are influenced more by the sense of place than space. In our study, the Room-type interface’s chair arrangement and classroom-like background transitioned a simple 2D space into a more engaging "place." Participants reported that Room-type discussions felt akin to real classroom interactions, unlike the Gallery-type.

Research on social presence suggests that a shared background can enhance the sense of unity among participants \cite{overview_presence}. In the Room condition, this background fosters a communal space, further strengthening the participants' feeling of togetherness. Additionally, MirrorBlender takes this concept further by incorporating screen sharing, offering participants control over their video window's position and opacity, enriching the virtual meeting experience \cite{gronbaek2021mirrorblender}.
% This functionality allows for unique interactions such as deictic pointing, resizing, and moving video windows to draw attention, 
\subsection{Design implications for video chat space}

\subsubsection{Use of Visual Cues for Video Chats with Different Group Structures}

Our comparison of Room-type and Gallery-type video chat interfaces reveals valuable design insights. Room-style configurations provide visual cues that enhance engagement and empathy through strategic chair placement, facilitating participant location, team awareness, and turn-taking. These arrangements demonstrate how users leverage scene properties for spatial recognition. However, Room configurations face a key limitation: visual cues remain static during chats. Preset arrangements based on anticipated team dynamics may misalign with actual interactions, potentially increasing participants' cognitive load \cite{sweller1998cognitive}. Future video chat interfaces should therefore incorporate adaptable visual cues that support specific tasks while avoiding cognitive overload from mismatched signals. Drawing from our findings on chair-based turn-taking, we propose that video chat designs should integrate flexible spatial cues. Gather Town \cite{gather_nodate} exemplifies this approach through game-like environments where users control movable avatars while maintaining video presence. This solution enables dynamic spatial reconfiguration to accommodate evolving team needs.

\subsubsection{Anxiety and the spotlight area in video chat}
Our analysis revealed participants' desire to "hide from the spotlight", highlighting how spatial metaphors of "seat" and "spotlight" relate to anxiety in video chats. In Gallery-type interfaces, "seats" refer to video positions in the grid layout (corner or center), while in Room-style interfaces, they correspond to chair locations in the scene. This spatial metaphor mirrors classroom seating choices, where spotlight areas vary by arrangement. In row-column layouts, corner positions parallel back-row seats, while central spots correspond to front-row positions. In circular arrangements, spotlight areas concentrate near the host. Generally, participants associate spotlight areas with heightened group attention, leading to anxiety from perceived observation. This understanding offers new perspectives on video chat interface design. Even with uniform video displays, certain positions may attract more attention and increase anxiety. While addressing participants' need to avoid spotlight areas remains important, simply hiding videos may not effectively reduce anxiety, as Miller's research suggests video feedback can protect against social anxiety \cite{k_miller_meeting_2021}. Future designs should explore spatial configurations that balance visibility with comfort.

\subsubsection{Enhancing Speaker Identification in Virtual Meetings}
In addition, even though Room-type introduces new methods for gaze tracking and referencing, some participants still find it difficult to discern who is speaking. They also feel that discussions in Room-type can be somewhat "chaotic," leading them to prefer just listening or relying on their memory of viewpoints. This encourages us to consider more friendly ways of tracking the speaker to know who is speaking.

In traditional Zoom meetings, enlarging the speaker's window helps in identifying the speaker effectively. However, this can also lead to discomfort due to being closely observed during video calls \cite{DeggesWhite2020ZoomAnxiety}. The design of Room-type can take ways like using the form of a stroke to make it clearer who the speaker is.
% Future research could also evaluate changing perspectives, such as using a first-person perspective~\cite{tang2023perspectives}. This would allow different attendees to view the room from their angle, and the gaze could be adjusted with the changing speakers without obstructing the view of the remaining participants. This approach could be somewhat similar to more immersive forms of remote meetings in VR~\cite{mcveigh2022beyond}.

\subsection{Limitations and Future Work}
We studied 8-9 person groups within specific age range (18-38) and mostly Chinese participants in this study. We also lacked quantitative survey data and within-subject comparisons. Moreover, we combined classroom scenes with seating arrangements in 2D, making it difficult to isolate effects. Finally, the video frame manipulation in Room condition may have influenced spatial perception.

Future work should explore demographics like elderly
people with different digital literacy and perceived telepresence than the younger generation \cite{liu2020you}, and gender-specific influence of the digital space. Most of the participants are from China, but different cultures across the globe might perceive space differently.

Recent work with procedural generation and Generative AI (GenAI) approaches~\cite{li_affecting_2024,yang_ai_2022} suggests that video conferencing may develop backgrounds that are automatically generated based on team collaboration~\cite{han_when_2024} or custom verbal specification~\cite{fu_being_2024,he_i_2025,zeng_ronaldos_2025}. Future work may explore how generative backgrounds affects spatial presence when they are applied to different meeting contexts. Further, interactive applications like games ~\cite{zhou_eternagram_2024,zhang_can_2025} and interactive art ~\cite{lc_together_2023,lc_human_2023} may take insight from this work in creating spatial coherence in the background environment with the interactions proposed.  

\bibliographystyle{splncs04}  % Specifies the style for the references
\bibliography{citations}     % Refers to your .bib file (change 'references' to your file name)

\end{document}